# Mutation Accumulation and the Catastrophic Senescence of Pacific Salmon.

T.J.P. Penna and S. Moss de Oliveira
*Instituto de Física, Universidade Federal Fluminense,*
*Av. Litorânea, s/n, 24210-340 Niterói, RJ, Brazil*
*e-mails: tjpp@if.uff.br and suzana@if.uff.br*

Dietrich Stauffer
*Institute for Theoretical Physics, Cologne University*
*D-50923 Köln, Germany*
*e-mail: stauffer@thp.uni-koeln.de*

(July 21, 1995)

The bit-string model of biological aging is used to simulate the catastrophic senescence of Pacific Salmon. We have shown that reproduction occuring only once and at a fixed age is the only ingredient needed to explain the catastrophic senescence according the mutation accumulation theory. Several results are presented, some of them with up to $10^8$ fishes, showing how the survival rates in catastrophic senescence are affected by changes in the parameters of the model.

Senescence, or aging, is a process occuring in all higher organisms and it is related to a decrease in functional abilities. Several factors seems to be important to aging: the environment, metabolism, genetic factors and so forth [1,2]. Senescence can be characterized as the decrease in the survival probabilities with the age. At least two theories for the senescence are based on evolution [3]: the antagonistic pleiotropy (an optimality theory based on life strategy of increasing fitness by increasing early performance at expense of late) and mutation accumulation (based on a greater mutation load on the later than the earlier ages). These theories, besides providing explanation for several aspects of senescence, also allow the use of methods of statistical physics (see ref. [4] for a review on earlier models). A dramatic manifestation of aging is the so-called catastrophic senescence of Pacific salmon, whereby they pass from sexual maturity to death in a few weeks. Semelparous individuals, which breed only once, usually present this feature [1–3], while for iteroparous individuals, which breed repeatedly, the senescence is more gradual. Jan [5] has shown how to introduce the catastrophic senescence in the Partridge-Barton model, but without taking into account explicitly the number of breeding attempts. In this work we show that, according to a model based on the mutation accumulation theory, this ingredient (semelparity) is the only responsible one for the catastrophic senescence.

The bit-string model of life history, recently introduced [6], makes use of a balance between hereditary mutations and evolutionary selection pressure to simulate aging in a population. In this model, each individual of an initial population $N(t = 0)$ is characterized by a string of 32 bits ("genome"), which contains the information when the effect of a mutation will be present during the life of the individuum. The time is a discrete variable running from 1 to 32 steps ("years"). If at time $(t = i)$ of the individuum lifetime, the $i$-th bit in the genome is set to one, it will suffer the effects of a deleterious mutation in that and all following years. At each year one bit of the genome is read, and the total number of mutations (bits 1) is computed; if it reaches a value greater than a threshold **T**, the individuum dies. At every year beyond the minimum reproduction age **R** the individuum produces **b** offsprings. The genome of each baby differs from that of the parent by one randomly selected bit, toggled at birth. This mutation can be regarded as hereditary variation arising from point mutation. In this paper, as well as in reference [9], only bad (deleterious) mutations are imposed at birth. If a bit=1 of the parent genome is chosen, it simply remains equal to one in the baby genome. On the other side, if a bit=0 of the parent genome is chosen, it is set equal to one in the baby genome.

The effect of food and space restriction is taken into account by an age-independent Verhulst factor, which gives to each individuum a probability $(1 - N(t)/N_{max})$ of staying alive; $N_{max}$ is typically ten times greater than the initial population $N(0)$, and represents the maximum possible size of the population. Results for the evolution of the whole population in time, and also the evolution in time according to different ages can be found in references [6–9]. In particular, the extension to 64-age intervals is presented in ref. [7]. There, the aging effects are easily noticed: there is always a greater number of youths than adults contributing to the whole population.

Some important features of this model are: a) It is based on mutation accumulation theory of aging ; b) It does not present the usual "mutational meltdown" resulting from accumulation of deleterious mutations (for more details of the mutational meltdown in this model see [7]); c) A large number of time intervals can be incorporated into the life history of an individual, instead of the only two (youth and adulthood) considered, for instance, in the Partridge-Barton model [3]; d) The parallelization of the algorithm on a multiple-data multiple-instruction computer with distributed memory is quite easy [8], allowing simulations with up to $10^8$ individuals. Differently from other problems in Monte Carlo simulations, as the Ising model for magnetic materials, we can work with sizes comparable to the real ones when studying population dynamics.

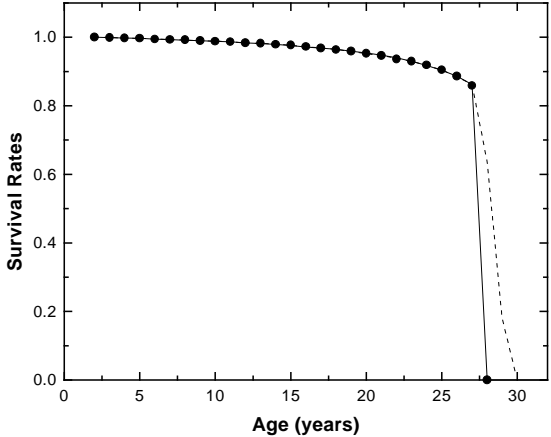

FIG. 1. Survival rates as a function of time in case of reproduction only at the reproduction age **R** = 27 (solid line), and in case of reproduction every year beyond a minimum age **R** > 26 (dashed line). The full circles correspond to a large-scale simulation, also in case of reproduction only at **R** = 27.

As mentioned before, the Pacific salmon suffers a sudden death just after reproduction - around 10 years. The oldest one found so far was 13 years old [10]. In order to check whether the mutation accummulation theory and the bit-string model are compatible with the catastrophic senescence phenomema, we proposed the following modification in the reproduction rule: Instead of producing **b** offsprings at each year beyond the reproduction age **R**, the fish produces **b** offsprings at, and only at, the reproduction age **R**. Figure 1 shows the survival rates as a function of time (years) for both reproduction rules. The survival rate is defined as $N_k(t)/N_{k-1}(t-1)$, that is, as the ratio between the population with age $k$ at time $t$ and the population with age $k-1$ at time $(t-1)$. We normalized the survival rates dividing them by the survival rate at age 1. To reduce the fluctuactions, we took the average population sizes in 300 steps, after 2700 steps. The reproduction age is **R** = 27 for the case of only one reproduction, and **R** > 26 for the other case. It is very difficult to compare one time step with real-life time scales, nevertheless, as we show below, these results are independent of this choice for the time scale. For both curves the other parameters are: $N(0) = 2 \times 10^5$, **T** = 1, and **b** = 30. The dots also presented in figure 1 correspond to a simulation performed with the same parameters, but starting with an initial population $N(0) = 6 \times 10^6$. No finite size effects for initial populations with at least $2 \times 10^5$ fishes are noticeable.

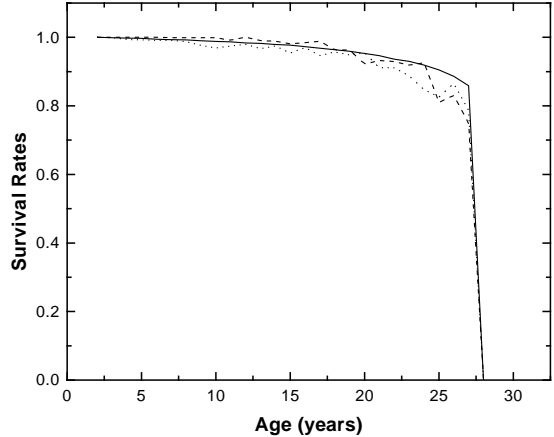

FIG. 2. Behavior of the survival rates in time for different values of the maximum number **T** of allowed mutations: **T** = 1 (solid line); **T** = 2 (dashed line); **T** = 3 (dotted line).

In figure 2 we present the results when the maximum number **T** of allowed mutations is varied. Only reproduction at a minimum age is considered, and the other parameters are the same as those in figure 1. It can be seen that a different value of **T** does not alter the survival rates although drastic changes occur in the original model. However, the final sizes of the populations are different for each case: the greater the value of **T**, the greater the final size of the population.

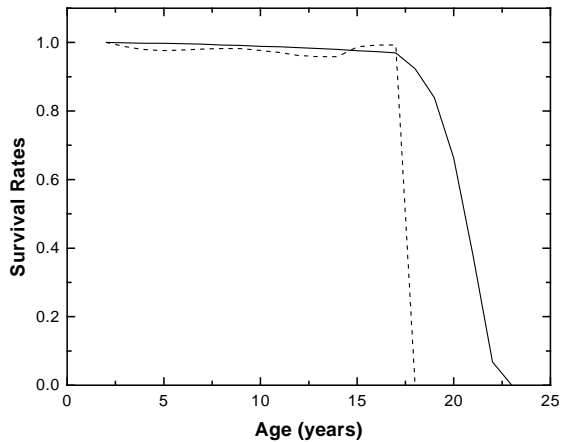

FIG. 3. The same as in fig.1, but for a different reproduction age: **R** = 17 (solid line), and **R** > 16 (dashed line).

Figure 3 is equivalent to figure 1, but for a different reproduction age value: **R** = 17 and **R** > 16 . It can be seen that the behavior for reproduction only at minimum age is the same independent of the value of **R**. Also the initial populations are smaller in this figure ($N(0) = 20,000$) than in fig. 1 ($N(0) = 200,000$) and, hence, fluctuations are evident as finite size effects. The difference between these two curves is bigger than the ones presented at fig.1 due to the effect of finite size of the bit-string (32 bits) is more effective for larger ages at reproduction.

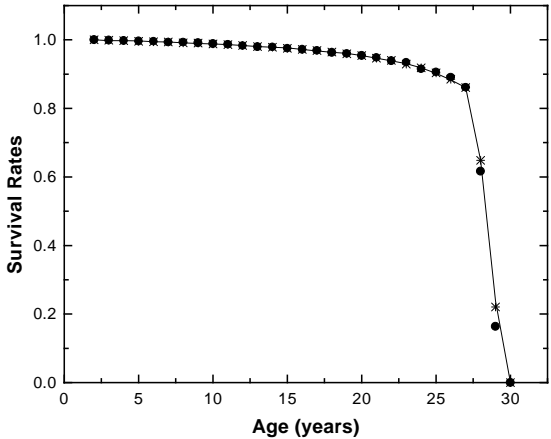

FIG. 4. Results for different birth rates in case of reproduction every year beyond the minimum age $\mathbf{R} > 26$: $\mathbf{b} = 10$ (solid line); $\mathbf{b} = 30$ ($\bullet$); $\mathbf{b} = 10$ ($*$), large-scale simulation.

In figure 4 we present the results for different birth rates in the case of reproduction every year beyond minimum age. Again it can be seen that the survival rates are independent of the value of the birth-rates, although the final sizes of the populations are not: $N_{b=30}(t = 3000) > N_{b=10}(t = 3000)$. The same effects are observed for the case of reproduction only at minimum age.

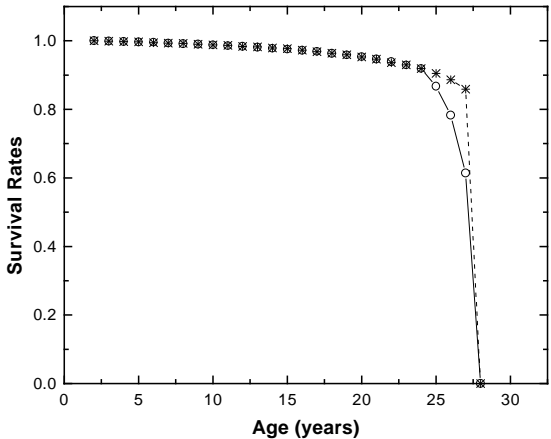

FIG. 5. Survival rates in time for the cases: reproduction at some age between $\mathbf{R} = 24$ and $\mathbf{R} = 27$ (o); the same reproduction age $\mathbf{R} = 27$ ($*$) for any fish.

As a last investigation we adopted the rule of reproducing only once, but at a minimum age randomly varying between 24 and 27 years. The result is shown in figure 5 and compared with reproduction only at $\mathbf{R} = 27$ for every fish. It can be noticed that the catastrophic senescence effect is more pronounced for only one reproduction age, however an additional senescence appears for the latter. The effect of variability on the age of reproduction is this model is presented at ref. [11]. Again the other parameters are the same as in fig.1. This result show us that both breeding once, and breeding for all fish at the same age, are responsible for catastrophic senescence.

In summary, we have shown that allowing each fish producing $\mathbf{b}$ offsprings only once, the catastrophic senescence of Pacific salmon can be nicely reproduced in the bit-string model for biological aging. Survival rates are unchanged for different values of the birth rate $\mathbf{b}$ and for different values of the maximum number of allowed mutations $\mathbf{T}$, although the final population sizes are sensible to these values. Also any value for the reproduction age $\mathbf{R}$ leads to the same result: accumulation of deleterious mutations extinguishes the post-reproductive population, because of the lack of selection pressure against these mutations. It is also shown that for the case in which the reproduction age is randomly chosen into a short range, the senescence effect is not so catastrophic as in the case that all fishes reproduce at the same reproduction age $\mathbf{R}$. It is well known that semelparous organisms show catastrophic senescence [3] but, as far as we know, this is the first model showing it explicitly.

### ACKNOWLEDGEMENTS

We thank Américo Bernardes for many discussions. This work is partially supported by the Brazilian agencies CNPq and FINEP. The simulations were carried out in IBM-RISC 6000 on the IF/UFF and on the Intel Paragon of KFA-Jülich.